\documentclass[twocolumn,abs,prb,showpacs]{revtex4-1}

\usepackage{graphicx}
\usepackage{dcolumn}
\usepackage{float}
\usepackage{amsmath}
\usepackage{bm}

\newcommand{\comment}[1]{}

\begin{document}


\title{Valley-Spin Polarization in the Magneto-Optical Response of Silicene and Other Similar 2D Crystals}

\author{C.J. Tabert$^{}$}
\author{E.J. Nicol$^{}$}
\affiliation{Department of Physics, University of Guelph,
Guelph, Ontario N1G 2W1 Canada} 
\affiliation{Guelph-Waterloo Physics Institute, University of Guelph, Guelph, Ontario N1G 2W1 Canada}
\date{\today}

\begin{abstract}
{We calculate the magneto-optical conductivity and electronic density of states for silicene, the silicon equivalent of graphene, and similar crystals such as germanene.  In the presence of a perpendicular magnetic field and electric field gating, we note that four spin- and valley-polarized levels can be seen in the density of states and transitions between these levels lead to similarly polarized absorption lines in the longitudinal, transverse Hall, and circularly polarized dynamic conductivity.  While previous spin and valley-polarization predicted for the conductivity is only present in the response to circularly polarized light, we show that distinct spin- and valley-polarization can also be seen in the longitudinal magneto-optical conductivity at experimentally attainable energies.  The frequency of the absorption lines may be tuned by the electric and magnetic field to onset in a range varying from THz to the infrared.  This potential to isolate charge carriers of definite spin and valley label may make silicene a promising candidate for spin- and valleytronic devices.}
\end{abstract}


\maketitle

\begin{figure}[h!]
\begin{center}
\includegraphics[width=0.9\linewidth]{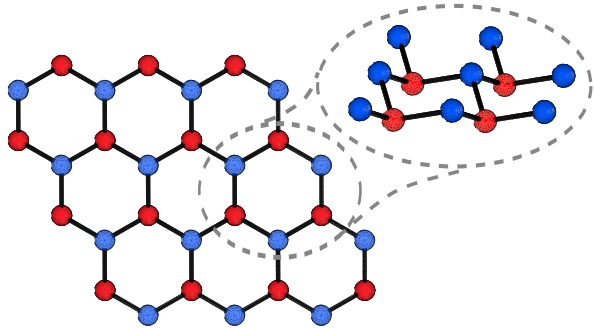}
\end{center}
\caption{\label{fig:Silicene}(Color online) The crystal structure\cite{Momma:2011} of silicene is based on the honeycomb lattice; the ionic size of the silicon atoms causes the A (red) and B (blue) sublattices to sit in separated parallel planes.
}
\end{figure}

\emph{Introduction:} Silicene, a monolayer of silicon atoms bonded together on a two-dimensional (2D) honeycomb lattice, has recently been synthesized\cite{Lalmi:2010, DePadova:2010, DePadova:2011, Vogt:2012, Lin:2012, Fleurence:2012} and begun to garner increased theoretical attention\cite{Guzman:2007, Topsakal:2010, Liu:2011, Liu:2011a, Ni:2012, Drummond:2012, Ezawa:2012, Ezawa:2012a, Ezawa:2012b, Stille:2012, Ezawa:2013, Tahir:2013} as it features a Dirac-like electron dispersion at the $K$ points of the Brillouin zone and promises to exhibit exciting properties beyond those present in graphene.  The similarities with graphene result from carbon and silicon residing in the same column on the chemical periodic table. The larger ionic size of silicon atoms causes the 2D lattice of silicene to be buckled\cite{Liu:2011, Liu:2011a, Drummond:2012} such that sites on the A and B sublattices sit in different vertical planes with a separation of $d\approx 0.46$ \AA \cite{Ni:2012, Drummond:2012} as illustrated in Fig.~\ref{fig:Silicene}. 
Because of the buckled lattice, an onsite potential difference ($\Delta_z=E_z d$) arises between the A and B sublattices when an electric field $E_z$ is applied perpendicular to the plane.  Silicene is also predicted to have a stronger intrinsic spin-orbit gap than is seen in graphene\cite{Konschuh:2010} with values (which can be increased under strain\cite{Liu:2011, Liu:2011a}) predicted to be $\Delta_{\rm so}\approx 1.55$ meV by density functional theory calculations\cite{Liu:2011, Liu:2011a, Drummond:2012} and is quoted as $\Delta_{\rm so}\approx 7.9$ meV in tight-binding calculations\cite{Liu:2011a}.  The larger spin-orbit interaction makes silicene susceptible to spin manipulation.  The resulting band gap near the two valleys $\bm{K}$ and $\bm{K^\prime}(\equiv -\bm{K})$ of the first Brillouin zone provides a ``mass" to the Dirac electrons that can be controlled by the strength of $\Delta_z$\cite{Ni:2012, Drummond:2012, Ezawa:2012a}.  Silicene has also been predicted\cite{Drummond:2012,Ezawa:2012a} to undergo a transition from a topological insulator (TI) (an insulator with a gapless spectrum of edge states\cite{Kane:2005, Hasan:2010}) to a band insulator (BI) as the strength of $\Delta_z$ becomes greater than $\Delta_{\rm so}$.  These qualities are also predicted for a monolayer of germanene which is isostructural to silicene but is expected to exhibit a much stronger spin-orbit band gap of $\Delta_{\rm so}\approx 24-93$ meV from first principles\cite{Liu:2011} and tight-binding calculations\cite{Liu:2011a}.  

When subjected to an external magnetic field, Landau levels (LLs) form in the electronic density of states and transitions between these levels result in absorption lines in the optical conductivity.  This has been discussed theoretically\cite{Gusynin:2007, Gusynin:2007b, Pound:2012} and confirmed experimentally\cite{Sadowski:2006, Sadowski:2007, Jiang:2007, Deacon:2007, Plochocka:2008, Orlita:2010, Henriksen:2010, Orlita:2011} for graphene.  In graphene, the $n=0$ LL is pinned at $\varepsilon=0$.\cite{Semenoff:1984, Gusynin:2007b, Gusynin:2007, Semenoff:2011, Pound:2012, Pound:2011, Pound:2011a}  When an excitonic gap is included\cite{Gusynin:2007b, Gusynin:2007}, the $n=0$ level splits between two distinct valley-polarized spin-degenerate energies but spin-polarized charge carriers are not expected in this case.  Conversely, due to spin-orbit coupling (SOC) and the response of silicene to an external perpendicular electric field, spin- and valley-polarized charge carriers appear due to the $n=0$ LL splitting between four spin and valley distinct energies.

In this Letter, we examine the effect of exposing silicene (or similar 2D crystals, such as germanene\cite{Liu:2011a}) to both external magnetic and electric fields with particular attention to the valley- and spin-polarized regions of the electronic density of states and dynamical conductivity.  This is of particular technological interest as four distinct valley- and spin-polarized currents can be generated through use of an in-plane electric field to generate a Hall current of spin- or valley-selected charges on the edges of the sample\cite{Xiao:2012,Tsai:2013}.  This ability to control the spin and valley index for use in such applications as data storage and data transmission is integral to valley-\cite{Xiao:2012, Zeng:2012, Mak:2012, Cao:2012} and spintronic\cite{Wolf:2001, Fabian:2007} devices.

\emph{Low Energy Hamiltonian:} It has been shown that the low-energy physics of silicene can be captured by the simple low-energy Hamiltonian\cite{Liu:2011a, Ezawa:2012, Ezawa:2012a, Ezawa:2012b}
\begin{equation}\label{Ham}
\hat{H}_\xi=\hbar v(\xi p_x\hat{\tau}_x+p_y\hat{\tau}_y)-\xi\frac{1}{2}\Delta_{\rm so}\sigma_z\tau_z+\frac{1}{2}\Delta_z\tau_z,
\end{equation}
where $\xi=\pm 1$ at the $K$ and $K^\prime$ points, respectively, $\sigma_z$ is the Pauli matrix associated with the electron's spin, $\bm{\tau}$ are the Pauli matrices associated with the sublattice pseudospin, $v\approx 5\times 10^5$m/s is the Fermi velocity of silicene and $p_x$ and $p_y$ are components of the momentum measured relative to the $K/K^\prime$ point.  The first term of the Hamiltonian is the familiar graphene-type Dirac Hamiltonian\cite{Neto:2009, Abergel:2010}.  The second term is the Kane-Mele term for intrinsic SOC\cite{Kane:2005} while the final term is associated with the onsite potential difference between the two sublattices that results from an external electric field\cite{Ezawa:2012, Ezawa:2012a, Ezawa:2012b}.  There is also a Rashba SOC but it may be ignored\cite{Ezawa:2012} for our purpose as it is a factor of 10 smaller than the intrinsic SOC.  

If we consider a perpendicular magnetic field and work in the Landau gauge, the magnetic vector potential, $\bm{A}=\nabla\times \bm{B}$, is written as $\bm{A}=(-By,0,0)$.  We make the Peierls substitution $\hat{p}_{i}\rightarrow\hat{p}_i+(e/c)\hat{A}_i$ in Eq.~\eqref{Ham} and obtain the low energy dispersion
\begin{equation}\label{Energy}
\varepsilon_n(\Delta_{\xi\sigma})=\left\lbrace\begin{array}{cc}
\text{sgn}(n)\sqrt{\Delta_{\xi\sigma}^2+2|n|v^2\hbar e B/c}, & n=\pm 1, \pm 2,...\\
-\xi\Delta_{\xi\sigma}, & n=0,
\end{array}\right.
\end{equation}
\begin{figure}
\begin{center}
\includegraphics[width=1.0\linewidth]{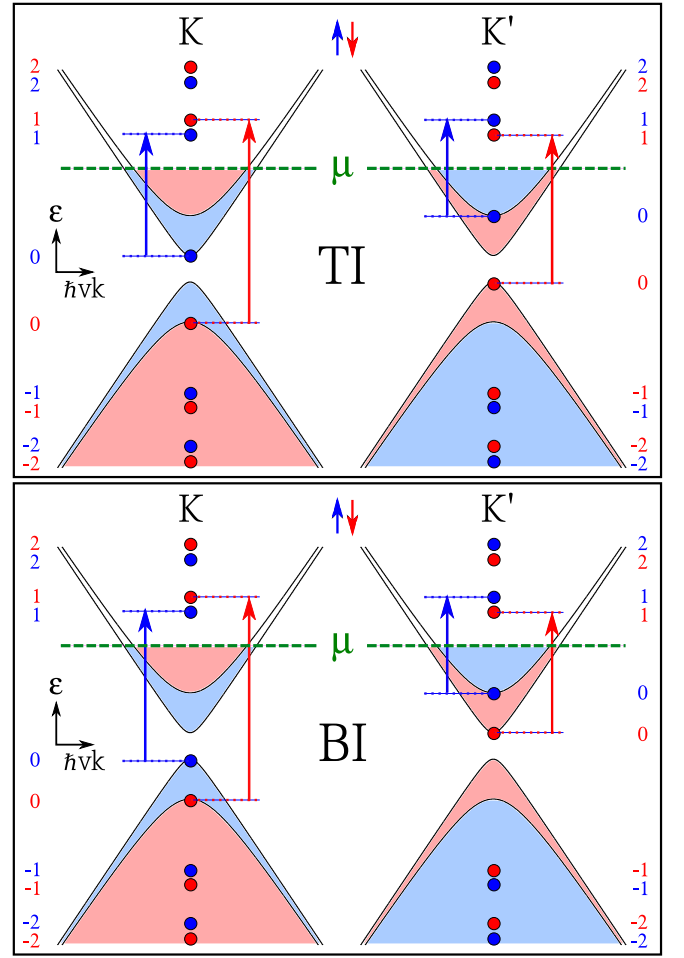}
\end{center}
\caption{\label{fig:Energy}(Color online) Schematic of the LLs (solid dots) compared to the band structure for $B=0$ (solid curves).  Blue/red represents spin up/down.  The numbers identify the LL index. Upper: TI regime ($\Delta_z<\Delta_{\rm so}$) at the $K$ (left) and $K^\prime$ (right) points of the Brillouin zone.  Lower: BI regime ($\Delta_z>\Delta_{\rm so}$).  A chemical potential $\mu$ (dashed green line) is shown that allows for spin-valley-polarized absorption lines in the magneto-optical conductivity due to the transitions identified with arrows.
}
\end{figure}
where $\Delta_{\xi\sigma}\equiv -\frac{1}{2}\sigma\xi\Delta_{\rm so}+\frac{1}{2}\Delta_z$, $\sigma=\pm 1$ for spin up and down, respectively, $e$ is the elementary charge and $B$ is the strength of the external magnetic field.  The Zeeman energy is small and can be ignored\cite{Tahir:2012, Ezawa:2012, Ezawa:2012a, Ezawa:2012b}.  In Fig.~\ref{fig:Energy} we show a schematic of the spin split bands at the $K$ and $K^\prime$ point for $B=0$ with the LLs for finite $B$ shown as dots.  The form of the zeroth LL has strong implications when considering the LLs in the TI ($\Delta_z<\Delta_{\rm so}$) vs. BI ($\Delta_z>\Delta_{\rm so}$) regimes which can be seen in the upper frame (TI) and lower frame (BI) of Fig.~\ref{fig:Energy}, respectively.  Given our expression for $\varepsilon_0(\Delta_{\xi\sigma})$, we can see that at both $K$ points in the TI regime, the $n=0$ spin up LL is at positive energy while the $n=0$ spin down level is at negative energy; refer to the upper frame of Fig.~\ref{fig:Energy}.  In the BI regime, $\Delta_{\rm so}$ is now greater than $\Delta_z$ and, therefore, the signs of $\varepsilon_0(\Delta_{K\uparrow})$ and $\varepsilon_0(\Delta_{K^\prime\downarrow})$ switch so that the $n=0$ spin up level is below zero at the $K$ point while the spin down $n=0$ level is above zero at the $K^\prime$ point as illustrated in the lower frame of Fig.~\ref{fig:Energy}.  This LL shift arises from the band inversion that results from the transition between the TI and BI regimes\cite{Ezawa:2012a}.

\begin{figure}
\begin{center}
\includegraphics[width=1.0\linewidth]{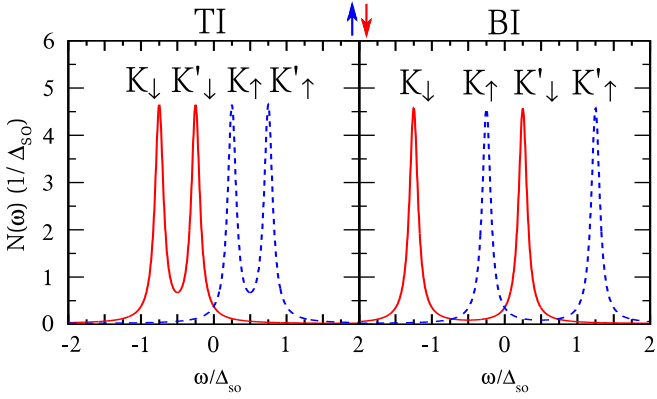}
\end{center}
\caption{\label{fig:DOS}(Color online) Spin-dependent density of states (red: spin down and blue: spin up). Left: in the TI regime ($\Delta_z=0.5\Delta_{\rm so}$).  Right: in the BI regime ($\Delta_z=1.5\Delta_{\rm so}$). Each plot is shown for $B/\Delta_{\rm so}^2$=0.0657T/meV$^2$ with a broadening parameter of $\eta=0.07\Delta_{\rm so}$.  There are four spin- and valley-polarized peaks at low energy with all higher peaks (not shown) being spin degenerate.
}
\end{figure}
\emph{Electronic Density of States:} Plots of the single-spin electronic density of states in the TI and BI regimes are shown in Fig.~\ref{fig:DOS}.  There are four spin-polarized peaks which onset at $\omega=-\Delta_{K\downarrow}$, $\Delta_{K^\prime\downarrow}$, $-\Delta_{K\uparrow}$ and $\Delta_{K^\prime\uparrow}$ corresponding to the four $n=0$ LLs.  The remaining higher energy peaks (not shown) are spin degenerate.  As scanning tunnelling spectroscopy experiments have been successful in observing LLs in graphene\cite{Li:2007, Miller:2009, Andrei:2012}, similar work on silicene should detect four distinct low energy spin- and valley-polarized levels.

\emph{Dynamical Conductivity:} Using the standard Kubo formula\cite{Nicol:2008, Tse:2011, Tabert:2012, Stille:2012, Tabert:2013}, we can derive expressions for the longitudinal and transverse Hall dynamical conductivities which also yield the familiar selection rules\cite{Gusynin:2007b, Gusynin:2007, Tse:2011} $n=|m|\pm 1$ for LL transitions.  For the absorptive part of the longitudinal conductivity we find
\begin{align}\label{Condxx}
&\frac{\text{Re}\sigma_{xx}(\Omega)}{\sigma_0}=\frac{2v^2\hbar e B}{c}\sum_{\sigma=\pm 1}\sum_{\xi=\pm 1}\sum_{n,m}\frac{\Theta(\mu-\varepsilon_m)\Theta(\varepsilon_n-\mu)}{\hbar\Omega}\notag\\
&\times\left[\left(A_{m}B_{n}\right)^2\delta_{n,|m|-\xi}
+\left(B_{m}A_{n}\right)^2\delta_{n,|m|+\xi}\right]\notag\\
&\times\delta(\hbar\Omega-(\varepsilon_n-\varepsilon_m)),
\end{align}
where $\sigma_0=e^2/(4\hbar)$ is the universal background conductivity of graphene.  For the absorptive imaginary part of the transverse Hall conductivity we find
\begin{align}\label{Condxy}
&\frac{\text{Im}\sigma_{xy}(\Omega)}{\sigma_0}=\frac{2v^2\hbar e B}{c}\sum_{\sigma=\pm 1}\sum_{\xi=\pm 1}\sum_{n,m}\xi\frac{\Theta(\mu-\varepsilon_m)\Theta(\varepsilon_n-\mu)}{\hbar\Omega}\notag\\
&\times\left[\left(A_{m}B_{n}\right)^2\delta_{n,|m|-\xi}
-\left(B_{m}A_{n}\right)^2\delta_{n,|m|+\xi}\right]\notag\\
&\times\delta(\hbar\Omega-(\varepsilon_n-\varepsilon_m)),
\end{align}
where
\begin{equation}\label{An}
 A_{n}=\left\lbrace\begin{array}{cc}
\displaystyle\frac{\text{sgn}(n)\sqrt{|\varepsilon_n|+\text{sgn}(n)\Delta_{\xi\sigma}}}{\sqrt{2|\varepsilon_n|}},& n\neq 0,\\\\
\displaystyle\frac{1-\xi}{2},& n=0,
\end{array}\right.
\end{equation}
and
\begin{equation}\label{Bn}
B_{n}=\left\lbrace\begin{array}{cc}
\displaystyle\frac{\sqrt{|\varepsilon_n|-\text{sgn}(n)\Delta_{\xi\sigma}}}{\sqrt{2|\varepsilon_n|}},& n\neq 0,\\\\
\displaystyle\frac{1+\xi}{2},& n=0.
\end{array}\right.
\end{equation}

\begin{figure}
\begin{center}
\includegraphics[width=1.0\linewidth]{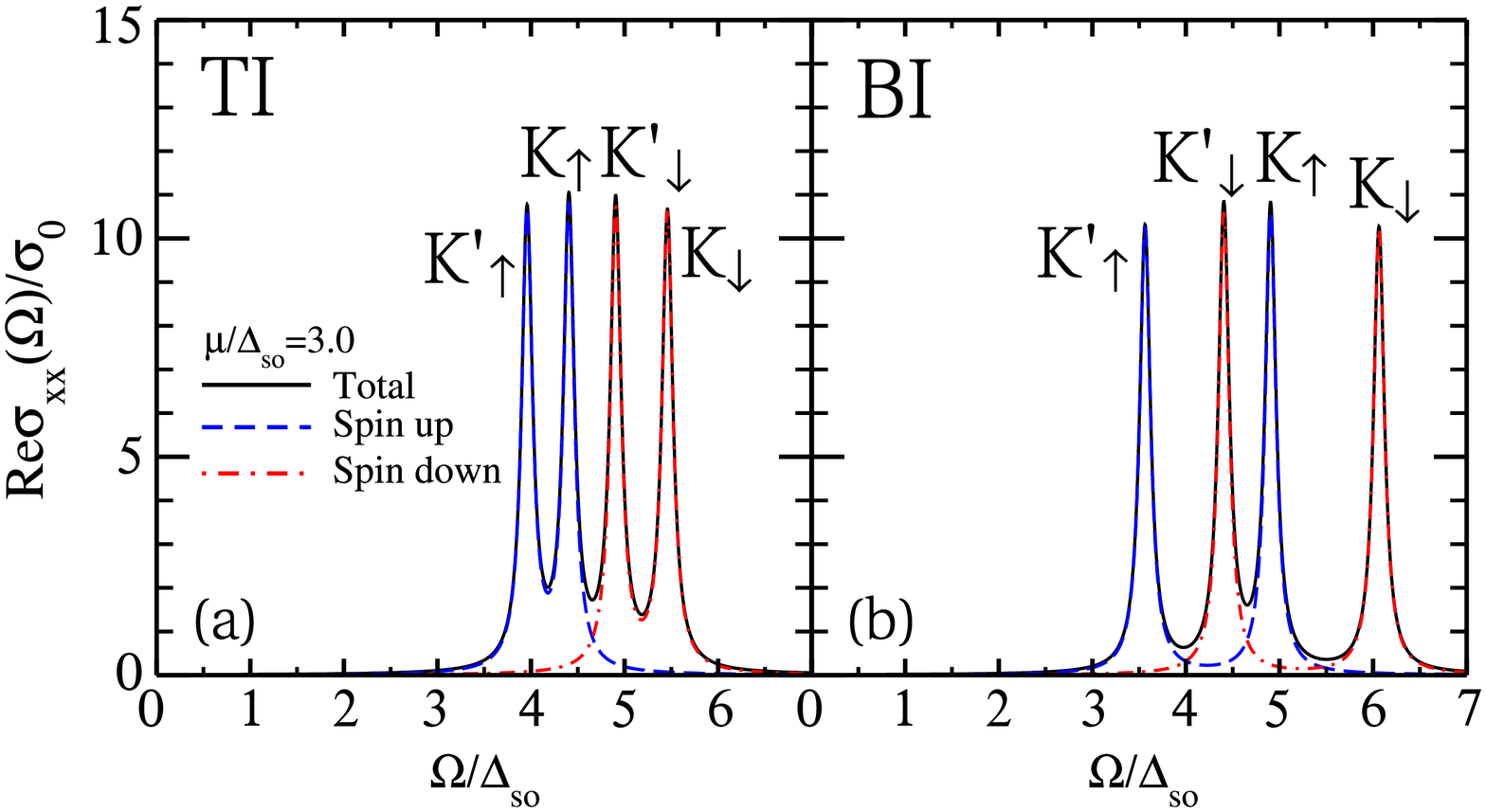}
\end{center}
\caption{\label{fig:Cond-spin}(Color online) Re$\sigma_{xx}(\Omega)$ in the (a) TI regime ($\Delta_z=0.5\Delta_{\rm so}$) and (b) BI regime ($\Delta_z=1.5\Delta_{\rm so}$) for $B/\Delta_{\rm so}^2$=0.0657T/meV$^2$ when the chemical potential lies between the $n=0$ and $n=1$ LLs.  The first four peaks are spin- and valley-polarized while higher features are not.
}
\end{figure}

The absorptive part of the longitudinal conductivity can be seen in Figs.~\ref{fig:Cond-spin}(a) and \ref{fig:Cond-spin}(b) for the TI and BI regimes, respectively.  The transverse Hall conductivity is not shown as in the region of spin and valley polarization it is simply the negative of the longitudinal conductivity.  We have used the same parameters as before and included a chemical potential of $\mu=3.0\Delta_{\rm so}$ so that the Fermi level lies above all the $n=0$ LLs and below all the $n=1$ levels. This is done so that all the $n=0$ LLs are occupied and, thus, transitions from $n=-1$ to $n=0$ are forbidden due to the Pauli exclusion principle while all $n=0$ to 1 transitions are permitted.  This will occur for any value of $\mu$ that is in this region, \emph{i.e.} $|\Delta_{K\uparrow}|<\mu<\varepsilon_1(\Delta_{K\uparrow})$.  The lower limit, $|\Delta_{K\uparrow}|$, is governed by the strength of the electric field while the upper bound, $\varepsilon_1(\Delta_{K\uparrow})$, can be raised by increasing $B$ and/or $E_z$. 

\begin{figure}
\begin{center}
\includegraphics[width=0.950\linewidth]{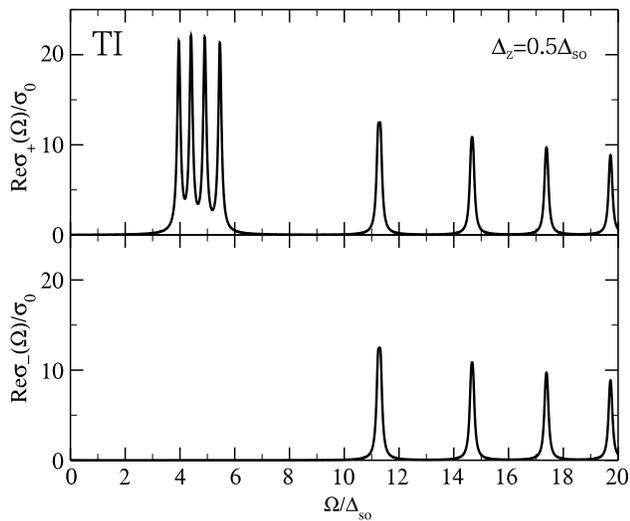}
\end{center}
\caption{\label{fig:Cond-circ} Conductivity for (upper) right-handed and (lower) left-handed polarized light.
}
\end{figure}

If the chemical potential is situated between $n=0$ and $|n|=1$, we see four absorptive peaks in the longitudinal conductivity which are spin and valley polarized.  These peaks result from single spin and valley transitions as illustrated by the arrows in Fig.~\ref{fig:Energy}.  While in the absence of a magnetic field spin-valley polarization has been predicted in the absorptive response to circularly polarized light\cite{Ezawa:2012b, Stille:2012}, the inclusion of a finite $B$ produces four strong, separated peaks of definite spin and valley label which can also be observed in the longitudinal response.  This is to be contrasted with the $B= 0$ response to circularly polarized light where predicted valley-spin polarization is not clearly separated but contains some admixture of spins and valleys\cite{Stille:2012, Zhou:2012}.  Similar to the $B=0$ result, spin and valley polarization can be seen in the response to circularly polarized light for $B\neq 0$.  Figure~\ref{fig:Cond-circ} shows the absorptive part of $\sigma_\pm(\Omega)$, where $\sigma_{\pm}(\Omega)=\sigma_{xx}(\Omega)\pm i\sigma_{xy}(\Omega)$ for right- and left-handed polarization, respectively\cite{Pound:2012}.  
There remains a spin- and valley-polarized quartet of peaks (of double weight) in response to right-handed polarization, while the quartet is lost in response to left-handed polarization and all that remains is the spin- and valley-degenerate response at higher energy.  In both the longitudinal and circularly polarized responses, the two middle peaks of the quartet display a band inversion\cite{Ezawa:2012a} with the transition from TI to BI regimes.

The magnetic field also allows for a tuning of the position of the valley-spin polarized peaks over a frequency range without having to adjust $E_z$.  The onset frequencies of these peaks is set by the energy difference between $n=0$ and 1 LLs, namely, $\Omega=\varepsilon_1(\Delta_{K^\prime\uparrow})-\Delta_{K^\prime\uparrow}$, $\varepsilon_1(\Delta_{K^\prime\downarrow})-\Delta_{K^\prime\downarrow}$, $\varepsilon_1(\Delta_{K\uparrow})-\Delta_{K\uparrow}$ and $\varepsilon_1(\Delta_{K\downarrow})-\Delta_{K\downarrow}$ associated with spin up and down at $K^\prime$ and spin up and down at $K$, respectively.  Thus, as $\varepsilon_1$ depends on $B$, increasing the magnetic field moves the quartet of peaks to higher energy.  While the separation between all four peaks is not fixed, the separation between the two middle peaks is fixed at $2|\Delta_{K\uparrow}|$ for $\Delta_z\neq 0$ (all levels are spin degenerate when $\Delta_z=0$) and as a consequence, when $\Delta_{z}=\Delta_{\rm so}$, only the outer peaks of the quartet remain spin-valley polarized.  This separation is the minimum gap between the $B=0$ bands and is only controlled by $E_z$ and the SOC.  

These polarized peaks should be visible in experiment as the magnetic and electric field values required to observe them are well within experimental limits.  Aside from $B$ and $E_z$, the determining factor in the onset frequency of the polarized response is the size of the spin orbit gap, $\Delta_{\rm so}$.  For $\Delta_{\rm so}=1.55$ meV, the curves shown here correspond to a $B$ of $\sim 0.15$ T and for $\Delta_{\rm so}=7.9$, $B\approx 4.1$ T.  In the former case, the splitting of peaks in the polarized quartet is $\sim 1.5$ meV, a resolution easily achieved in broadband optics.  In this case, the quartet sits in the range of $1-2$ THz where broad band experiments can be done\cite{Mori:2008}.  If $B=1$ T, such peaks will shift to the far infrared in the range of $17-20$ meV. If instead $\Delta_{\rm so}$ is as large as 8 meV, the splitting of the quartet will be about 3-5 meV for $B=1$ T and the quartet will fall in the range of $13-25$ meV.  Conductivity experiments on graphene have spanned the range from THz to eV\cite{Sadowski:2006, Sadowski:2007, Jiang:2007, Deacon:2007, Henriksen:2010, Crassee:2011} with B$\sim 0$ T to 18 T\cite{Sadowski:2006, Sadowski:2007, Jiang:2007, Deacon:2007, Henriksen:2010, Crassee:2011} and resolutions of order less than a meV; therefore, experiments on silicene should observe this polarized behaviour.  Experiments may also provide a measure of $\Delta_{\rm so}$, for instance, by examining when the two middle polarized peaks become superimposed at the transition between the TI and BI regimes ($\Delta_z=\Delta_{\rm so}$).

While electron-electron interactions have been a source of much discussion in graphene\cite{Kotov:2012, Goerbig:2011}, experimental magneto-optics in graphene has been well-described by the single-particle picture up to very large magnetic fields of $\sim$ 16T\cite{Sadowski:2006, Sadowski:2007, Jiang:2007, Deacon:2007, Plochocka:2008, Orlita:2010, Goerbig:2011}, hence for modest magnetic fields ($\sim$1T), electron-electron interactions should not modify our main results.  In addition, our calculations have been based on the assumption of freestanding silicene or silicene on an insulating substrate where the main effect is to provide charge doping, for example, by back gating.  Currently silicene is fabricated on Ag substrates, but efforts are underway to find alternative substrates, most recently\cite{Fleurence:2012}, ZrB$_2$. As in graphene, the choice of insulating substrate is not expected to affect qualitatively the results presented here.

\emph{Summary:} We have predicted the presence of a quartet of spin- and valley-polarized peaks in the low energy electronic density of states which result in spin- and valley-polarized absorption lines in the magneto-optical longitudinal and transverse Hall dynamical conductivities of silicene as well as in the response to circularly polarized light.  These absorption peaks arise from transitions between $|n|=0$ and 1 LLs at both valleys and are only present when the chemical potential lies between the $|n|=0$ and 1 LLs at the $K(K^\prime)$ point.  The energy at which these peaks occur can be controlled by the strength of the magnetic and electric fields.  Furthermore, the spin and valley labels of the two middle peaks switch as the system transitions from a TI to a BI.  As the strength of the spin- and valley-polarized response is predicted to be comparable to what has been observed for graphene and to onset at experimentally attainable energies, it should be easily visible in experiment.   While spin-valley polarization in the absence of a magnetic field has been discussed\cite{Ezawa:2012b,Stille:2012}, it can only be isolated with circularly polarized light and may contain some admixture of spin and valley.  With a finite $B$, such spin-valley polarization is robust, easily separated out and can be observed without circular polarization.  This may prove useful in both spintronic and valleytronic technologies as an electronic current of definite spin and valley label can potentially be isolated.  These predictions should hold for other 2D crystals with a band gap that can be tuned by an external electric field ($\emph{i.e.}$ germanene).

We thank J. P. Carbotte for helpful discussions.  This work has been supported by the Natural Science and Engineering Research Council of Canada.


\bibliographystyle{apsrev4-1}
\bibliography{llsi-lett}

\end{document}